\documentstyle[preprint,aps,epsfig]{revtex}
\begin{document}
%
%
\newcommand{\C}{\mbox{$^{12}$C}}
\newcommand{\Cg}{\mbox{$^{12}$C$_{\rm g.s.}$}}
\newcommand{\N}{\mbox{$^{12}$N}}
\newcommand{\Ng}{\mbox{$^{12}$N$_{\rm g.s.}$}}
\newcommand{\num}{\mbox{$\nu_{\mu}$}}
\newcommand{\numb}{\mbox{$\bar{\nu}_{\mu}$}}
\newcommand{\nub}{\mbox{$\bar{\nu}$}}
\newcommand{\nue}{\mbox{$\nu_e$}}
\newcommand{\nueb}{\mbox{$\bar{\nu}_{e}$}}
\newcommand{\mum}{\mbox{$\mu^-$}}
\newcommand{\mup}{\mbox{$\mu^+$}}
\newcommand{\el}{\mbox{$e^-$}}
\newcommand{\pos}{\mbox{$e^+$}}
\newcommand{\pim}{\mbox{$\pi^-$}}
\newcommand{\pip}{\mbox{$\pi^+$}}
\newcommand{\pipd}{\mbox{\pip\,$\rightarrow$\,\mup\,+\,\num}}
\newcommand{\mupd}{\mbox{\mup\,$\rightarrow$\,\pos\,+\,\nue\,+\,\numb}}
\newcommand{\CC}{\mbox{\C\,(\,\nue\,,\,\el\,)\,\N }}
\newcommand{\excl}{\mbox{\C\,(\,\nue\,,\,\el\,)\,\N$_{\rm g.s.}$}}
\newcommand{\msigma}{\mbox{$\langle\,\sigma\,\rangle$}}
\newcommand{\numbnueb}{\mbox{\numb\,$\rightarrow\,$\nueb}}
\newcommand{\gemu}{\mbox{$g^{\gamma}_{\epsilon \mu}$}}
\newcommand{\scaten}{\mbox{$\mid g^{S}_{RL}\,+\,2\,g^{T}_{RL}\mid$}}
\newcommand{\sig}{\mbox{$\langle\sigma\rangle$}} 
\newcommand{\Ql}{\mbox{$Q_L^{\nu}$}}
\newcommand{\wl}{\mbox{$\omega_{L}$}}
\newcommand{\Em}{\mbox{$E_{\rm max} = 52.8$~MeV}}
\newcommand{\NIM}{Nucl. Instrum. and Methods Phys. Res., Sect. }
\newcommand{\PR}{Phys. Rev. }
\newcommand{\PRL}{Phys. Rev. Lett. }
\newcommand{\PL}{Phys. Lett. }
\title{Measurement of the Energy Spectrum of \nue\ from Muon Decay 
and Implications for the Lorentz Structure of the Weak Interaction}

\author{B.~Armbruster,$^1$ I.M.~Blair,$^2$ B.A.~Bodmann,$^3$ N.E.~Booth,$^4$ 
G.~Drexlin,$^1$ V.~Eberhard,$^1$ J.A.~Edgington,$^2$ C.~Eichner,$^5$ 
K.~Eitel,$^1$ E.~Finckh,$^3$ H.~Gemmeke,$^1$ J.~H\"o\ss l,$^3$ T.~Jannakos,$^1$ 
P.~J\"unger,$^3$ M.~Kleifges,$^1$ J.~Kleinfeller,$^1$ W.~Kretschmer,$^3$
R.~Maschuw,$^5$ C.~Oehler,$^1$ P.~Plischke,$^1$ J.~Rapp,$^1$ C.~Ruf,$^5$ 
M.~Steidl,$^1$ O.~Stumm,$^3$ J.~Wolf,$^{1*}$ B.~Zeitnitz$^1$}

\address{$^1$ Institut f\"ur Kernphysik I, Forschungszentrum Karlsruhe,
  Institut f\"ur experimentelle Kernphysik, Universit\"at Karlsruhe,
  Postfach 3640, D-76021 Karlsruhe, Germany}
\address{$^2$ Physics Department, Queen Mary and Westfield College,
  Mile End Road, London E1 4NS, United Kingdom}
\address{$^3$ Physikalisches Institut, Universit\"at Erlangen-N\"urnberg,
  Erwin-Rommel-Stra\ss e 1, D-91058 Erlangen, Germany}
\address{$^4$ Department of Physics, University of Oxford,
  Keble Road, Oxford OX1 3RH, United Kingdom}
\address{$^5$\it Institut f\"ur Strahlen- und Kernphysik, Universit\"at Bonn,
  Nu\ss allee 14--16, D-53115 Bonn, Germany}
\address{$^*$ Present address: Department of Physics and Astronomy,
  University of Alabama, Tuscaloosa AL 35487}
\maketitle
\begin{abstract}
The KARMEN experiment uses the reaction \excl\ to measure the energy 
distribution of \nue\ emitted in muon decay at rest \mupd . The \nue\ analog 
\wl\ of the famous Michel parameter $\rho$ has been derived from a 
maximum-likelihood analysis of events near the kinematic end point, 
\Em. The result, \mbox{\wl $= (2.7^{+3.8}_{-3.3}\pm 3.1) \times 10^{-2}$}, 
is in good agreement with the standard model prediction $\wl= 0$. We deduce a 
90\% confidence upper limit of $\wl \le 0.113$, which corresponds to a limit 
of \scaten $ \le 0.78$ on the interference term between scalar and tensor 
coupling constants. 
\end{abstract}
\vspace{1cm}
Experimental results from nuclear $\beta$ decay and muon decay form the basis
of the \mbox{V-A} hypothesis, which is an essential feature of the standard 
model (SM) of electroweak interactions. The rate of muon decay, the purely 
leptonic process \mupd, has been used to determine the universal Fermi coupling 
constant $G_{F}$. Precise measurements of the shape of the \pos\ energy 
spectrum, the decay asymmetry between the \mup\ spin and \pos\ momentum, and 
the polarisation vector of the \pos\ have led to bounds on the scalar, vector 
and tensor coupling constants, which form the Lorentz structure of the charged 
weak interaction. These results combined with the inverse process 
\mbox{\num$\:$ +$\:$\el$\:\rightarrow$ \mum + \nue} underpin the SM
assumption of lepton number conservation, the \mbox{V-A} interaction and
universality \cite{fet0}. All experiments up to now support the \mbox{V-A} 
structure of the weak interaction; however, substantial non- (\mbox{V-A}) 
components are not ruled out.

Complementary to these experiments, which are all based on observation of the 
charged leptons only, the Karlsruhe Rutherford 
Medium Energy Neutrino experiment (KARMEN) 
determines the energy spectrum of the \nue\ emitted in the decay \mupd\ of
unpolarized muons to draw conclusions on the Lorentz structure. In the 
well-known case of \pos\ spectroscopy, it is the Michel parameter $\rho$ 
which governs the shape of the \pos\ energy spectrum. In an analogous way, the 
shape of the \nue\ energy spectrum is determined by the parameter \wl, which 
also depends on vector, scalar, and tensor components of the weak interaction, 
but in a different combination. In the SM all non- (\mbox{V-A}) 
components vanish, and \wl\ is predicted to be $0$. Thus an upper limit on 
\wl\ derived from the analysis of the \nue\ energy spectrum provides new 
limits on nonstandard couplings.

All features of muon decay are most generally described by a local, 
derivative-free, lepton-number-conserving, four-lepton point interaction 
with the matrix element given by \cite{fet1} 
\begin{equation} 
M = \frac{4}{\sqrt{2}} G_{F} \sum_{\gamma = S,V,T \atop
\epsilon,\mu = R, L} \gemu \langle
\bar{e}_{\epsilon} |\Gamma^{\gamma}| (\nue)_n \rangle \langle
(\numb)_m|\Gamma_{\gamma}|\mu_{\mu} \rangle.
\end{equation}
The index $\gamma$ labels the type of interaction $\Gamma$ (S = 4-scalar, 
V = 4-vector, T = 4-tensor) and the indices $\epsilon$ and $\mu$ indicate the 
chirality (L = left-, R = right-handed) of electron and muon spinors, 
respectively. In this re\-presentation the chirality of the neutrino $n$ or $m$
 is fixed to be equal to that of the associated charged lepton for the 
V interaction, but opposite for the S and T interactions. As $G_{F}$ sets the 
absolute strength of the interaction, the ten coupling constants \gemu\ are 
dimensionless complex quantities normalized by
\begin{equation} \label{norm}
3 |g^T_{RL}|^2+ 3 |g^T_{LR}|^2+ \sum_{\epsilon,\mu = R, L} (\frac{1}{4}
|g^S_{\epsilon \mu}|^2+ |g^V_{\epsilon \mu}|^2 ) = 1
\end{equation}
with $g^{T}_{RR} = g^{T}_{LL} \equiv 0$. In the SM, muon decay is 
a pure V interaction mediated between left-handed particles, so all coupling 
constants  vanish except $g^{V}_{LL} \equiv 1$. Although this representation 
is elegant from the theoretical point of view, the individual coupling 
constants cannot be determined directly by experiment. However, the measurable 
parameters ($\rho$, $\eta$, \wl, etc.) are expressable as positive 
semidefinite bilinear combinations of \gemu\ from which upper or 
lower limits for the coupling constants can be derived.

The possibility of measuring \wl\ with the KARMEN experiment was first pointed
out by Fetscher \cite{fet2}. More recently Greub et al.\ \cite{fet3} have 
calculated the spectrum of left-handed \nue\ including radiative corrections 
and effects of finite lepton masses. Taking significant terms only, the
spectrum $dN_L/dx$ can be described by
\begin{equation}\label{nuspec}
\frac{dN_L}{dx} = \frac{G^2_F m^5_{\mu}}{16\pi^3} \, \Ql \, \{G_0(x) + 
G_1(x) + \wl G_2(x)\} 
\end{equation} 
where $m_{\mu}$ is the muon mass, $x=2 E_{\nu}/m_{\mu}$ is the reduced neutrino
energy, and \Ql\ denotes the probability of emission of a left-handed \nue. 
The function $G_0(x)$ describes the pure V-A interaction, $G_1(x)$ takes into 
account radiative corrections, and $\wl G_2(x)$ includes the effect of scalar 
and tensor components according to 
\begin{equation} \label{ome}
\wl = \frac{3}{4}\cdot
\frac{|g^S_{RR}|^2 + 4|g^V_{LR}|^2 + |g^S_{RL} + 2 g^T_{RL}|^2}
{|g^S_{RL}|^2 + |g^S_{RR}|^2 +4|g^V_{LL}|^2+4|g^V_{LR}|^2+12|g^T_{RL}|^2} .
\end{equation}

\begin{figure}
  \centerline{\epsfig{figure=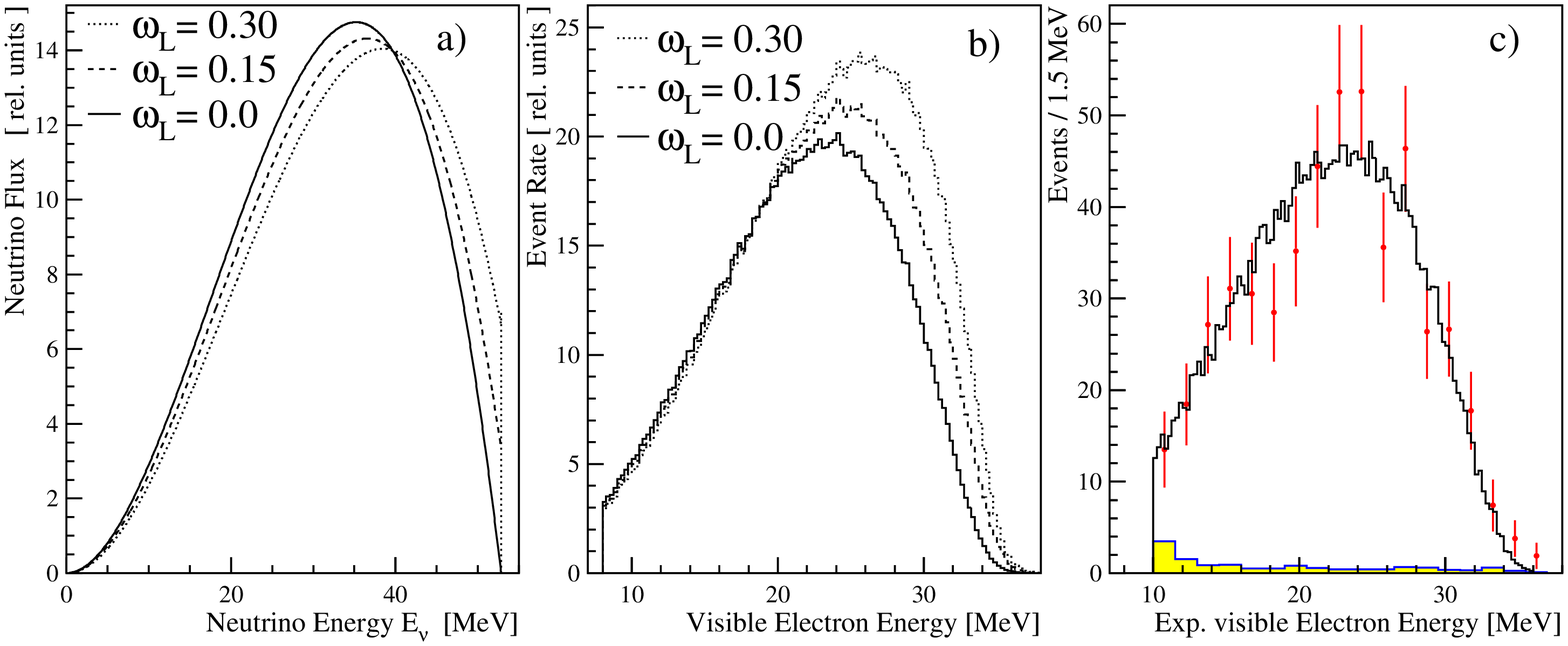,width=17.8cm}}
  \caption{Influence of different values of $\wl=0.0$, 0.15, 0.3 on (a) the 
\nue\ energy spectrum in \mup\ decay and on (b) the visible electron energy 
spectrum measured with the reaction \excl\ (c) Experimental electron energy 
distribution together with MC expectation (solid line) and the subtracted 
background (shaded).}
  \label{wtheory}
\end{figure}

The calculated \nue\ energy spectra for different values of \wl\ are shown in
Fig.\ \ref{wtheory}(a). Momentum conservation in the decay fixes the emission 
direction of \nue\ near the kinematic end point to be opposite to that of the 
positron and the \numb. Together with angular momentum conservation this 
implies suppression of emission of left-handed \nue\ in the case of vector 
coupling, while all other couplings enhance the decay rate at the end point. 
The total decay rate, and therefore the integral neutrino flux, is unchanged by 
nonstandard interactions.

The KARMEN experiment uses the pulsed spallation neutron facility ISIS at the 
Rutherford Appleton Laboratory to investigate neutrinos from \mup\ decay. The 
800~MeV proton beam from ISIS is stopped in a Ta-D$_2$O target producing 
neutrons and pions. All charged pions are stopped inside the target within 
$10^{-10}$~s, the \pim\ being absorbed by the heavy target material while 
the \pip/\mup\ decay chain \pipd,~ \mupd\ produces an intense burst of \num, 
\nue, and \numb, emitted isotropically with equal intensity. Since both 
\pip\ and \mup\  decay at rest, the energy spectra of the neutrinos 
are well defined. The \pip\ decay produces monoenergetic \num\ with 
$E_{\nu_\mu} = 29.8$~MeV; the \nue\ and \numb\ from the \mup\ decay have 
continuous energy distributions up to \Em. The time structure 
of ISIS --- two 100~ns wide proton bunches 324~ns apart and recurring at 
50~Hz --- determines the production time of the different $\nu$ flavors: the 
short \pip\ lifetime ($\tau_{\pi} = 26$~ns) leads to two \num\ pulses within 
the first 500~ns after beam-on target. These pulses are well separated in time 
from the production of \nue\ and \numb, which follow with the much longer 
lifetime of the \mup\ ($\tau_{\mup} = 2.2~\mu$s). This leads to a suppression 
factor of about $10^{4}$ for cosmic-ray background.
  
The neutrinos are detected in a segmented 56~ton liquid scintillation 
calorimeter consisting of 512 optical modules, each with a length of 3.53~m 
and a cross section of $18 \times18$~cm$^2$ \cite{kar}. The detector is an 
almost completely (96\%) active calorimeter optimized for the measurement of 
electrons around 30~MeV and achieves 
resolutions of $\sigma(E)/E = 11.5\%/\sqrt{E\mbox{(MeV)}}$ for energy, 
and $\sigma(X) \approx 7$~cm for position measurement. A 7000~ton shielding 
steel blockhouse together with two layers of active veto counters suppresses 
beam-correlated spallation neutrons and cosmic-ray muons.

The signature that unambiguously identifies a \nue\ is a delayed coincidence 
consisting of an electron from the charged current reaction \excl\ in the time 
window of \nue~production followed by a positron from the subsequent 
$\beta$ decay of \Ng\ ($\tau = 15.9$~ms) at the same location in the detector. 
Each event fully contained within the central detector with time
0.6--9.6~$\mu$s after beam-on target and energy 10--36~MeV is identified 
as electron, provided it is followed by a positron event within 
0.5--36~ms with energy 3.5--16.5~MeV. We demand the sequence to be detected 
in the same or adjacent module within a distance $\Delta X \le 35$~cm along the 
module axis. Cuts used to reduce cosmic background are the same as used in 
previous data evaluations \cite{cc1}. In data accumulated between June 
1990 and December 1995 --- corresponding to 9122~C of protons or 
$2.51\times 10^{21}$ \mup~decays in the ISIS target --- we find 513 \el /\pos\ 
sequences. Subtracting $13.3\pm0.8$ background events and taking into account 
an overall detection efficiency $\epsilon =32.8\%$, the flux-averaged cross 
section is 
\begin{equation}\label{exp1}
 \sig_{\rm exp} = (9.4\pm0.4{\rm (stat)}\pm0.8{\rm (sys)})
\times 10^{-42}\mbox{~cm}^2\mbox{.} 
\end{equation}
This is in good agreement with different theoretical calculations of 
$\sig_{\rm th}$ in the range of \mbox{(9.1--9.4)$\times 10^{-42}$ cm$^2$}
 \cite{theo,Don}. 

As the recoil energy of the \N\ nucleus is negligible, the \nue\ energy 
$E_{\nu}$ is determined from the measurement of the electron energy $E_e$ via
the kinematic relation $E_{\nu} = E_e + Q$, where $Q=17.3$~MeV is the $Q$ value
of the detection reaction. The energy dependence of the cross section is
dominated by the phase-space factor $(E_{\nu}-Q)^2$. Therefore, a low rate of
additional \nue\ at the kinematic end point \Em\ due to nonstandard
couplings is translated to the observation of a significantly higher rate of
electrons and thus to a distortion of the visible energy spectrum of Fig.\ 
\ref{wtheory}(b).

\begin{figure}
  \centerline{\epsfig{figure=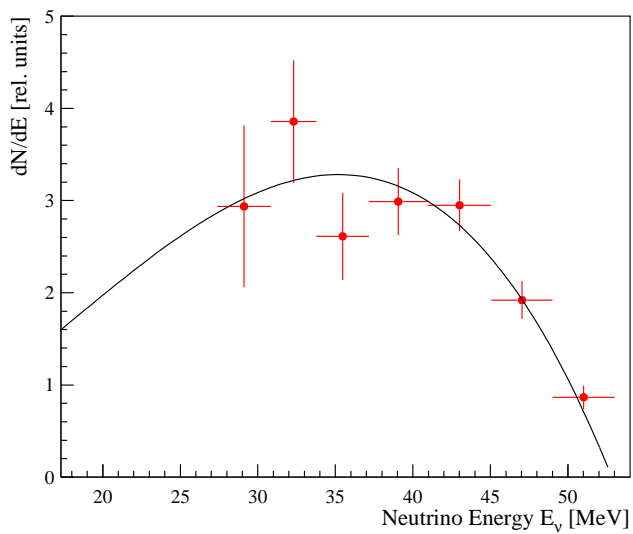,width=8.6cm}}
  \caption{Energy spectrum of \nue\ from \mup\ decay determined by an 
unfolding method compared with the standard model expectation (solid line).} 
  \label{fold}
\end{figure}

The KARMEN calorimeter allows a precise measurement of the energy $E_{e}$ 
[see Fig.\ \ref{wtheory}(c)]. The energy
spectrum of \nue\ from \mup\ decay is then determined in two steps. First, we
apply the procedure of regularized unfolding described by Blobel \cite{blo} to
derive the true electron energy. This method takes into account the detector 
response and minimizes inherent instabilities (oscillating solutions) by 
demanding a priori a certain degree of smoothness of the true electron 
distribution depending on statistical accuracy. The \nue\ energy distribution 
is then calculated from the number of primary electrons, within a given 
interval $\Delta E$ from the unfolding procedure, divided by the corresponding 
mean cross section. This yields a \nue\ energy spectrum with seven data points 
as shown in Fig.\ \ref{fold} and compared with the V-A expectation. This
represents the first measurement of the neutrino energy spectrum from muon
decay in addition to the well-known \pos\ spectrum.

Because of the strong energy dependence of the detection cross section the most 
detailed information on \wl\ and \Ql\ is obtained from the experimental 
electron spectrum of Fig.\ \ref{wtheory}(c). The analysis is done by two 
independent methods: (1) the investigation of the measured decay rate on the 
basis of the flux-averaged cross section, and (2) the analysis of the spectral 
shape with a maximum likelihood (ML) method. 

\begin{figure}
  \centerline{\epsfig{figure=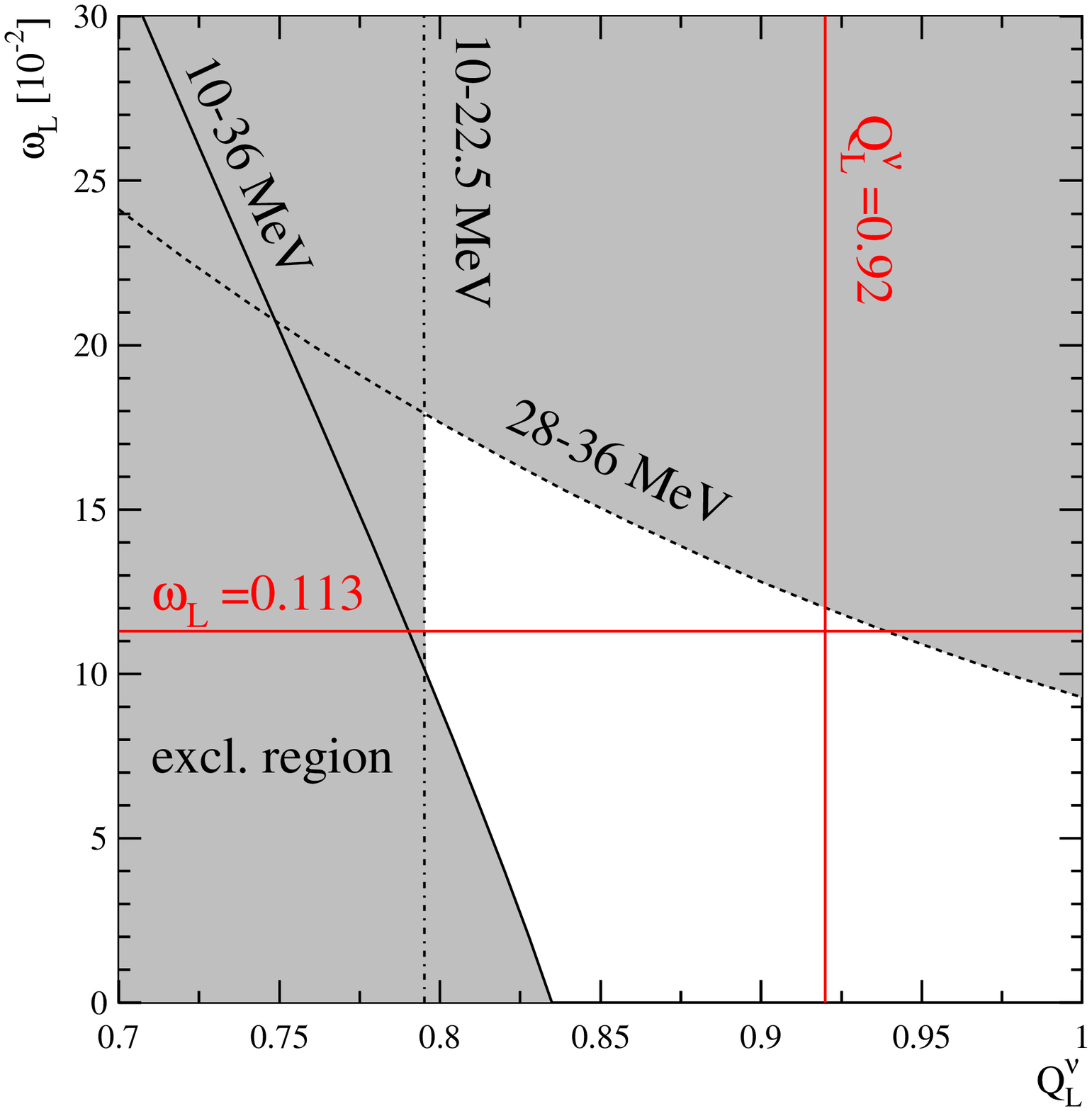,width=8.6cm}}
  \caption{The \Ql-\wl\ parameter space: The shaded regions are excluded at 
90\% confidence from the different analyses of the absolute flux in several 
energy ranges. The horizontal line is the result of the spectral shape analysis 
$\wl \le 0.113$ at 90\% confidence. The vertical line is the current best
limit $\Ql \ge 0.92$.}     
  \label{wabs}
\end{figure}

As can be seen from Fig.\  \ref{wtheory}(b), $\wl > 0$ would result in 
additional \excl\ events; on the other hand, $\Ql < 1$ would reduce the number 
of events [see Eq.\ (\ref{nuspec})]. In order to find allowed regions in the 
\Ql-\wl\ parameter space, we compared measured and expected flux averaged 
cross sections. As theoretical cross section $\sig_{\rm th}$ with a realistic 
estimate of the systematic error we use $\sig_{\rm th} = 
(9.2\pm0.5)\times10^{-42}$~cm$^2$. The experimental cross section 
is taken from Eq.\ (\ref{exp1}) with statistical and systematic error added 
quadratically. The probability distribution of the ratio 
\begin{equation}
R(\Ql,\wl) = \frac{\sig_{\rm exp}}{\sig_{\rm th}} = \Ql (1+ S\cdot \wl) = 
\frac{9.4\pm0.9}{9.2\pm0.5}
\end{equation}
incorporates a flux decrease by right handed \nue\ through \Ql\ as well as an 
increase by nonzero \wl\ values; $S$ is the ratio of additional events in 
case of $\wl = 1$ relative to the expectation for $\wl=0$. We have sampled the 
probability density function of the ratio $R$ from Gaussian distributions of 
$\sig_{\rm exp}$ and $\sig_{\rm th}$ for 3 different energy ranges: (a) the 
range 
10--36~MeV with the highest statistical accuracy, but only moderate sensitivity 
$S=0.81$, (b) the range 28--36~MeV, where with $S=3.48$ we are very sensitive 
to \wl, and (c) the range 10--22.5~MeV, where the expected event number is 
almost independent of \wl\ [$S=0.002$, see Fig.\ \ref{wtheory}(b)]. From range 
(c) we deduce a lower limit $\Ql \ge 0.796$. The shaded parameter space shown 
in Fig.\ \ref{wabs} combines regions excluded at 90\% confidence level of all 3 
energy ranges. From inverse muon decay experiments it is known that 
$\Ql > 0.92$ \cite{fet4,PDG}. Including this information in our analysis of 
range (b) restricts the allowed area and sets a 90\% confidence upper limit 
$\wl \le 0.12$.

In the second method we determine \wl\ by analyzing the shape of the visible 
electron spectrum independent of \Ql. In order to increase the energy 
resolution and to reduce the background level we applied more stringent cuts on 
the electron position along the module axis $|X| \le 150$~cm and on the 
electron time 0.6--7.2~$\mu$s. These cuts reduce the background to only 6.0 
events in a sample of 441~events, thus nearly doubling the 
signal-to-background ratio.

The theoretical \nue\ energy spectrum of Eq.\ \ref{nuspec} was converted into a
visible electron spectrum using the energy-dependent $\sigma(E_{\nu})$ taken
from \cite{Don} folded with the detector response by a MC calculation. The ML
procedure was carried out on an event-by-event basis for several fit intervals
all of which gave results compatible with $\wl = 0$ within a $1\sigma$-error.
The net result is
\begin{equation}
\wl = (2.7^{+3.8}_{-3.3}\mbox{(stat)}\pm 3.1{\mbox{(sys)}})
\times 10^{-2} .
\end{equation}
Including the systematic error (energy shift of 0.25~MeV or 0.7\% scaling
error) we find, with the most conservative Bayesian 
approach, a 90\%~confidence upper limit $\wl \le 0.113$. This excludes the
region above the horizontal line in Fig.\ \ref{wabs}. Combining Eq.\ 
(\ref{norm}) and Eq.\ (\ref{ome}) the following relation between the shape 
parameter \wl\ and nonstandard couplings is \cite{fet4,not1}
\begin{equation}
|g^S_{RL} + 2 g^T_{RL}|  \le  \sqrt{\frac{16}{3} \wl} \mbox{.} 
\end{equation}
The limit on \wl\ thus results in an upper limit of $\scaten \le 0.78$ for the 
interference term of scalar and tensor amplitudes. 

In conclusion, the KARMEN experiment finds no evidence for nonstandard 
coupling constants in \mup\ decay at rest, either by a determination of the 
absolute \nue\ flux or by analysis of the spectral shape. Our analysis excludes 
most of the \Ql-\wl\ parameter space and yields for the first time an upper 
limit on the neutrino Michel parameter \wl.
 
During 1996 the experiment was upgraded by an additional active veto
counter in order to increase the sensitivity of the search for neutrino
oscillations in the channel \numbnueb \cite{upg}. Since 1997 KARMEN has been 
taking data again. Up to the end of 1999 we expect about 400 further charged
current events, which will reduce the statistical error by about a factor of 
1.4. Considering also a reduction of the systematic error, this may result in 
a limit competitive with the present best limit $\scaten \le 0.45$ deduced from 
measurements of the positron polarization \cite{fet4}.

We gratefully acknowledge the financial support from the German 
Bundesministerium f\"ur Bildung, Wissenschaft, Forschung und Technologie 
(BMBF), the Particle Physics and Astronomy Research Council (PPARC), and the 
Central Laboratory of the Research Council (CLRC). In particular, we thank 
W.~Fetscher for numerous discussions.


\begin{references}
\bibitem{not1} Limits on $|g^S_{RR}|$ and $|g^V_{LR}|$ are possible as well, 
                but are not as stringent as in other experiments.
\end{references}
\end{document}